\documentclass[11pt]{aa}
\usepackage[english]{babel}
\usepackage{graphicx}
\usepackage{longtable}

\onecolumn
\newcommand{\ab}{Astrophys. Bull. }
\newcommand{\arep}{Astron. Rep. }
\newcommand{\alet}{Astron. Let. }

\newcommand{\mnras}{Mon. Not. R. Astron. Soc. }
\newcommand{\apj}{Astrophys. J. }

\newcommand{\aj}{Astron. J. }
\newcommand{\aaa}{Astron and Astrophys.}
\newcommand{\aas}{Astron and Astrophys. Suppl.}
\newcommand{\pasp}{Publ. Astron. Soc. Pasif. }

\begin{document}

\title{Spectral variability of the IR--source IRAS\,01005+7910 optical component}

\author{V.G.~Klochkova, E.L.~Chentsov, V.E.~Panchuk, E.G.~Sendzikas, M.V.~Yushkin}
\institute{Special Astrophysical Observatory RAS, Nizhnij Arkhyz,  369167 Russia}

\date{\today} 

\abstract{Highly--resolution optical spectra of the  optical component of the IR--source IRAS\,01005+7910 are used to 
determine the spectral type of its central star, B1.5$\,\pm\,$0.3, identify the spectral features, and analyze their 
profile and radial velocity variations. The systemic velocity \mbox {$V_{\rm sys} = -50.5$~km\,s$^{-1}$} is determined
from the positions of the symmetric and stable profiles of the forbidden [N\,I], [N\,II], [O\,I], [S\,II], and 
[Fe\,II] emission lines. The presence of the [N\,II] and [S\,II] forbidden emissions indicates the onset of 
the ionization of the circumstellar envelope and the fact that the star is very close to undergoing
the planetary nebula stage. The broad range of heliocentric radial velocity V$_r$ estimates based on the core lines, 
which amounts to about 34\,km\,s$^{-1}$, is partly due to the deformations of the profiles
caused by variable emissions. The variations of the V$_r$ in the line wings are smaller, about 23\,km\,s$^{-1}$, 
and may be due to pulsations and/or hidden binarity of the star. The deformations of the profiles of complex absorption--emission 
lines may result from variations of their absorption components caused by the variations of the  geometry and kinematics 
in the wind base. The H$\alpha$ lines exhibit P\,Cyg\,III~type wind profiles. Deviations of the wind from spherical 
symmetry are shown to be small. The relatively low wind velocity (27--74\,km\,s$^{-1}$ from different
observations) and the strong intensity of the red emission (it exceeds the continuum level by up to a factor of seven) are
typical for hypergiants rather than the classical supergiants. IRAS\,01005 is an example of spectral mimicry of a low mass
post-AGB star masquerading as a massive hypergiant. 
}

\titlerunning{\it Spectral variability of the IR-source IRAS\,01005+7910}
\authorrunning{\it Klochkova et al.}

\maketitle

\section{Introduction}

The IR--source IRAS\,01005+7910 (hereafter referred to as IRAS\,01005) is located high above the Galactic 
plane at a latitude of  $b = 16\fdg6$. Its optical counterpart is a peculiar B--type supergiant, B\,=\,11$\fm5$, V\,=\,11$\fm2$. 
The position of the source in the  IRAS--color diagram is consistent with it undergoing the stage of 
a protoplanetary nebula (PPN). According to the chronological sequence of Lewis~[\cite{Lewis}], the absence of 
OH and CO emissions~[\cite{Likkel1989,Likkel1991,Nyman,Omont,Hu94}] indicates that the object is close to the planetary 
nebula stage (PN). Klochkova et al.~[\cite{01005}] obtained essential results for the central star of IRAS\,01005. 
The above authors determined the fundamental parameters of the supergiant (T$_{\rm eff}$\,=\,21500\,K, surface gravity $\log g =3.0$, 
metallicity [Fe/H]\,=$-0.31$, and the abundances of a number of chemical elements) and confirmed its post--AGB status. 
Another important result of the above study is the discovery of carbon overabundance ($\mbox{C/O}>1$) in the atmosphere of 
the central star.

The IR spectrum of the circumstellar envelope of IRAS\,01005 exhibits spectral features of the carbon-containing 
fullerene molecule (C$_{60}$)~[\cite{Zhang,Iglesias}]. IRAS\,01005 is the hottest post--AGB star among those whose 
spectra exhibit the so far unidentified features at  3.3 and 3.4~$\mu$m~[\cite{Hrivnak2007}]. The high Galactic 
latitude of IRAS\,01005 combined with its low metallicity indicates that the star belongs to the old population of the Galaxy.

Several teams studied the photometric variability of  IRAS\,01005. Hrivnak et al.~[\cite{Hrivnak2010}] noted 
variations of the object on a time scale of less than several days which is too short for PPNe. 
Arkhipova et al.~[\cite{Arkhip2013}] performed long-term $UBV$ photometric monitoring of several hot
PPNe including  IRAS\,01005 and found these objects to exhibit rapid irregular light variations. 
Based on these observations of the low-amplitude PPN light variations, Arkhipova et al.~[\cite{Arkhip2013}] 
concluded that the parameters of the stellar winds of these stars show variations and/or
micropulsations with characteristic periods of several hours.

The optical spectrum of IRAS\,01005 has so far not been studied in sufficient detail. Based on 
low--resolution observations, Hu~[\cite{Hu}] determined its spectral type to be Sp\,=\,B2\,Ie and pointed 
out the P\,Cyg profile of the H$\alpha$ line in the spectrum of IRAS\,01005. He found no differences 
between the spectra of the star taken ten years apart. However, Klochkova et al.~[\cite{01005}] identified
spectral features in the spectra taken with echelle spectrographs of the 6--m telescope of the Special 
Astrophysical Observatory and found variations in the spectrum of the  central star. The spectrum contains  
C\,II, O\,II, N\,II, Al\,III, and Si\,III absorption lines, the Mg\,II $\lambda\,4481$\,\AA\ line, and
emission features identified with Si\,II and the forbidden~[Fe\,II] lines. The profiles of the hydrogen Balmer
lines, the resonance Na\,I doublet, the He\,I and Fe\,III lines contain both the emission and absorption components. 
The optical spectrum was found to show substantial variations: the neutral helium lines vary from direct 
to inverse  P\,Cyg profile on time scales from several days to several months. The lines of the Na\,I
resonance doublet contain five absorption components with the velocities of  $-11$, $-28$, $-52$, $-65$, and
$-73$~km\,s$^{-1}$, and the absorption profile of the $\lambda\,5890$ line is superimposed onto the high-velocity
emission component whose width coincides with that of the emission components of the hydrogen lines. 
The observed variability of the spectrum provides a stimulus for further investigation of the
optical component of the IR source IRAS\,01005.

In this paper we report the results of long-term monitoring of IRAS\,01005 carried out with the aim to study 
the variability of the optical spectrum. Note that unlike our previous study~[\cite{01005}], here we use only the
high-resolution spectroscopic data. Section~\ref{method} briefly describes the technique
of observations and data reduction and Section~\ref{results} presents the main results.

\begin{table*}
\caption{Spectroscopic data and mean heliocentric radial velocities, V$_r$}
\medskip
\begin{tabular}{ c|  c|  c|  c  c| c| c| c| c| r| c }
\hline
Data& $\Delta\lambda$, & $S/N$&  \multicolumn{8}{c}{V$_r$,~km\,s$^{-1}$}\\
\cline{4-11}
    &                  &     &\multicolumn{2}{c|}{Main}           & Forbid  &\multicolumn{2}{c|}{Absorptions} &Em.$/$abs. & \multicolumn{2}{c}{Abs.$/$em.} \\
    &                  &     &\multicolumn{2}{c|}{components}     & emis  &\multicolumn{2}{c|}{}               &Fe\,III, (He\,I) & \multicolumn{2}{c}{} \\
\cline{4-11}
\cline{4-11}
&  && \multicolumn{2}{c|}{Na\,I, (Ca\,II)} & & wings&  cores &  &  \multicolumn{1}{c|}{H$\beta$}&  \multicolumn{1}{c}{H$\alpha$} \\
\hline
 {(1)}        & (2) & \multicolumn{1}{c|}{(3)} & \multicolumn{2}{c|}{(4)} & \multicolumn{1}{c|}{(5)} & (6) & (7) & \multicolumn{1}{c|}{(8)} & \multicolumn{1}{c|}{(9)} & \multicolumn{1}{c}{(10)} \\
\hline
25.01.2002  &  354--500 & 40  &  $(-71:$  & $-12:) $ & $-50: $ &  $-45$  &   $-40:$  &   $-78/-37$     & $-77/-31$  &            \\
04.02.2002  &  460--607 & 60  &  $-72.5$  & $-10.0 $ & $-51.0$ &  $-46$  &   $-44 $  &   $-44/$        & $-75/-35$  &            \\
22.11.2002  &  538--685 & 50  &  $-72.5$  & $-9.9  $ & $-49.5$ &  $-50$  &   $-45 $  &                 &            &  $-82/-32$ \\
25.11.2002  &  538--685 & 30  &  $-72.3$  & $-10.3 $ & $-50: $ &  $-43$  &   $-30 $  &                 &            &  $-83/-31$ \\
27.11.2002  &  452--600 & 35  &  $-71.5$  & $-9.3  $ & $-51.1$ &  $-44$  &   $-41 $  &   $-45/$        & $-86/-23$  &            \\
02.12.2002  &  452--600 & 45  &  $-72.5$  & $-10.2 $ & $-51.0$ &  $-42$  &   $-29 $  &   $-56/-26$     & $-87/-24$  &            \\
03.12.2002  &  452--600 & 30  &  $-71.9$  & $-10.3 $ & $-49.2$ &  $-42$  &   $-26 $  &   $-58/-25$     & $-86/-27$  &            \\
19.12.2002  &  452--600 & 50  &  $-72.7$  & $-10.6 $ & $-50.4$ &  $-47$  &   $-37 $  &   $-70:/-34:$   & $-80/-35$  &            \\
23.02.2003  &  516--666 & 55  &  $-72.5$  & $-9.2  $ & $-51: $ &  $-50$  &   $-40 $  &   $-80:/-38:$   &            & $-101/-27$ \\
13.04.2003  &  528--676 & 80  &  $-72.3$  & $-10.0 $ & $-48: $ &  $-43$  &   $-33 $  &                 &            & $-100/-19$ \\
15.11.2003  &  352--500 &100  &  $(-70.5$ & $-10.8)$ & $-49.6$ &  $-32$  &   $-19 $  &   $(-70/-29)$   &  $-81/-24$ &            \\
10.01.2004  &  528--676 & 50  &  $-72.2$  & $-10.3 $ & $-49.8$ &  $-50$  &   $-51 $  &                 &            & $-124/-32$ \\
09.03.2004  &  528--676 & 55  &  $-72.3$  & $-9.5  $ & $-50.6$ &  $-46$  &   $-37 $  &                 &            & $-110/-25$ \\
28.08.2004  &  528--676 & 55  &  $-72.5$  & $-10.1 $ & $-51: $ &  $-48$  &   $-45 $  &                 &            & $-86/-27$ \\
18.01.2005  &  528--676 &110  &  $-72.8$  & $-10.2 $ & $-50.9$ &  $-55$  &   $-53 $  &                 &            & $-77/-28$ \\
13.11.2005  &  456--601 &110  &  $-72.3$  & $-10.1 $ & $-50.6$ &  $-45$  &   $-37 $  &   $-64/-29$     &  $-75/-33$ &            \\
15.11.2005  &  528--678 &120  &  $-72.7$  & $-10.1 $ & $-51.0$ &  $-43$  &   $-30 $  &   $(-70:/-35:)$ &            & $-82/-28$ \\
09.12.2006  &  447--594 &130  &  $-72.3$  & $-10.6 $ & $-49.7$ &  $-46$  &   $-25 $  &   $-55/-22$     &  $-80/-29$ &            \\
03.11.2008  &  446--593 &160  &  $-71.8$  & $-10.0 $ & $-50.3$ &  $-41$  &   $-24 $  &   $-53/-18$     &  $-85/-30$ &            \\
05.11.2008  &  446--593 &160  &  $-72.3$  & $-10.5 $ & $-50.7$ &  $-44$  &   $-30 $  &   $-54/-20$     &  $-90/-27$ &            \\
20.11.2010  &  397--545 & 75  &           & $-50.5$  &  $-46$  &  $-29 $ &   $-55/-20$&  $-82/-30$     &            \\
29.05.2013  &  391--680 & 70  &  $-72.5$  & $-9.8$   & $-50.5$ &  $-45$  &   $-40 $  &   $-72/-33$     &  $-77/-30$ &  $-87/-25$ \\
21.08.2013  &  391--680 & 80  &  $-71.8$  & $-10.0$  & $-51.1$ &  $-40$  &   $-27 $  &   $-60/-22$     &  $/-26$    &  $-103/-27$\\
\hline
\end{tabular}
\label{spectra}
\end{table*}

\section{Observational data and their reduction}\label{method}

In this study we use twenty three high-resolution spectra (R\,=60\,000) taken with the NES echelle
spectrograph~[\cite{nes1,nes2}] of the 6--m telescope of the Special Astrophysical Observatory during the
period from 2002 to 2013. It is important for spectral variability studies to use only the spectra 
taken with the same spectrograph, albeit in different wavelength intervals. Unfortunately, the
spectra also differ by their exposure levels (signal-to-noise ratios $S/N$). The dates, wavelength intervals, 
and maximum $S/N$ ratios are listed in the first, second, and third columns of Table\,\ref{spectra} respectively. 
To reduce the spectra, we used the DECH program~[\cite{gala}], which, in particular, can smooth  $r(\lambda)$ curves 
without appreciable loss of resolution and measure radial velocities from paricular features of complex lines, 
which are typical of our spectra, by superimposing the direct and mirrored images of their profiles.

All the radial velocities mentioned in this paper are heliocentric. The laboratory wavelengths used to determine these
velocities are listed in Table\,\ref{lines}. Most of them are adopted from the NIST database\footnote{\tt
http://www.nist.gov/pml/data/asd.cfm}  and were validated by the spectra of the stars 10\,Lac (O9\,V) and 
$\iota$\,Her (B3\,III) with narrow lines taken with the same spectrograph as the one used
to observe  IRAS\,01005. In a number of cases (He\,I triplets, C\,II and Mg\,II doublets, etc.) we use the effective
wavelengths. We indicate the number of the multiplet in parentheses both in the text and in Table\,\ref{lines}.

\begin{figure}[]
 \vspace{2mm}
\includegraphics[angle=0, width=0.6\textwidth, bb=25 5 635 525, clip]{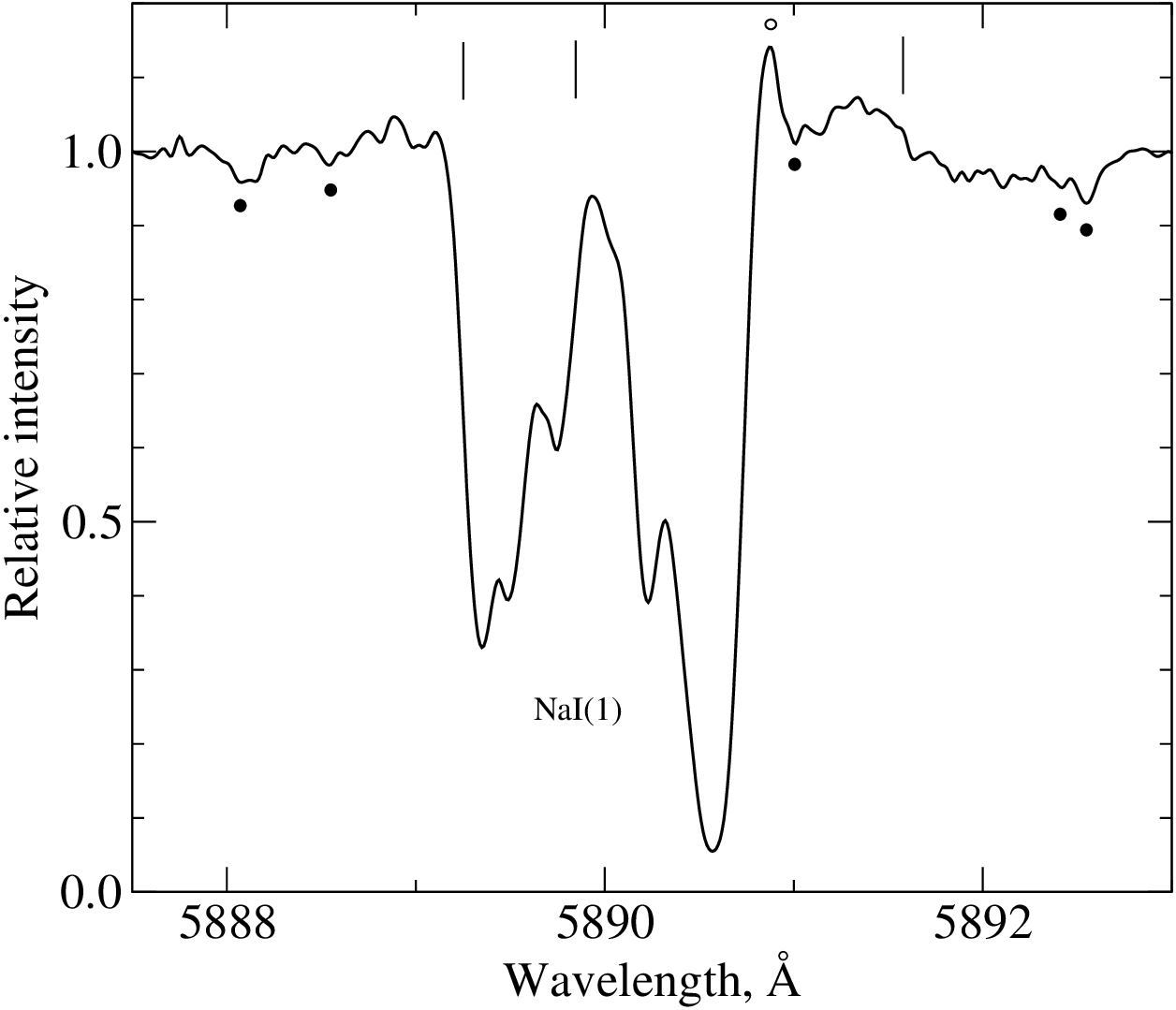}
\caption{A fragment of the spectrum of IRAS\,01005 taken on November 13, 2005 with the multicomponent
         interstellar absorption Na\,I\,(1) 5889.95~\AA{} and the emission-absorption line C\,II\,(5) 5891.59~\AA{}. 
         The horizontal axis gives the scale of laboratory wavelengths of photospheric absorptions. The positions of 
         the lines of the C\,II\,(5) multiplet are marked by vertical dashes, those of telluric H$_2$O
         absorptions---by the filled circles, and those of the telluric Na\,I emission---by the open circle.} 
\label{fig1}
\end{figure}

If telluric lines were present in the spectrum we used these to control and correct the inferred radial velocities. 
We additionally controlled the stability of the radial  velocity system by interstellar lines, mostly those of  Na\,I\,(1). 
We took into account the distortions due to the blending of  these lines by telluric  H$_2$O absorptions and Na\,I 
emissions, and, in the case of the D2 line, also by the  C\,II\,(5) triplet. All these distortions vary in time. 
Figure\,\ref{fig1} illustrates their presence in the spectrum of November 13, 2005 and the complex structure of the  
D2~Na\,I\,(1)\,5890\,\AA{} line profile. The heliocentric radial velocities of its five main components averaged 
over all our data are equal to  $-72.5$, $-65.3$, $-52.2$, $-27.7$, and $-10.2$~km\,s$^{-1}$.
The  velocities of the two outermost (the deepest) components are listed in Table\,\ref{spectra} (column~4). 
Judging by these velocities, the systematic errors of the radial velocities listed in other columns of Table\,\ref{spectra}
do not exceed 1~km\,s$^{-1}$.

\section{Main  results}\label{results}

\subsection{Line-profile and radial velocity variations}

Klochkova et al.~[\cite{01005}] showed that the profiles of most of the lines vary both in time and within the same
spectrum, from line to line. The profiles and their presumable formation regions can, to a first approximation, be subdivided
into three types:
\begin{list}{}{}
 \item $\bullet$ narrow emissions---the tenuous extended envelope;
 \item $\bullet$ emission-absorption profiles---the transition region from the photosphere to the envelope and, in particular, 
  stellar wind;
 \item $\bullet$ absorptions---the photosphere.
\end{list}

\begin{figure}
\includegraphics[angle=0, width=1.0\textwidth, bb=3 260 737 525, clip]{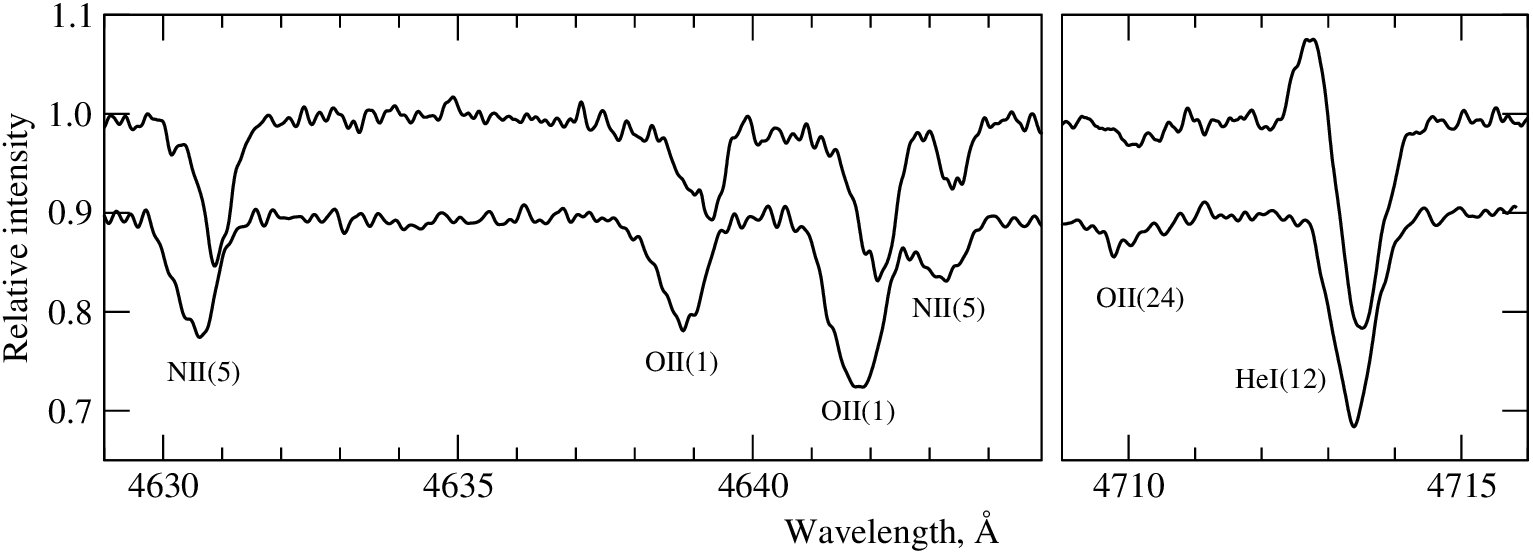}
\caption{Variations of the shapes and positions of the profiles in the spectrum of IRAS\,01005. Fragments 
         of the spectra taken on November 15, 2003  (top) and November 13, 2005 (shifted downward by 0.1) are 
         smoothed by seven points. The horizontal axis is marked by the laboratory wavelengths of photospheric 
         absorptions in the spectrum of November 13, 2005.}
\label{fig2}
\end{figure}

\begin{figure}
 \vspace{1mm}
\includegraphics[angle=0, width=0.65\columnwidth]{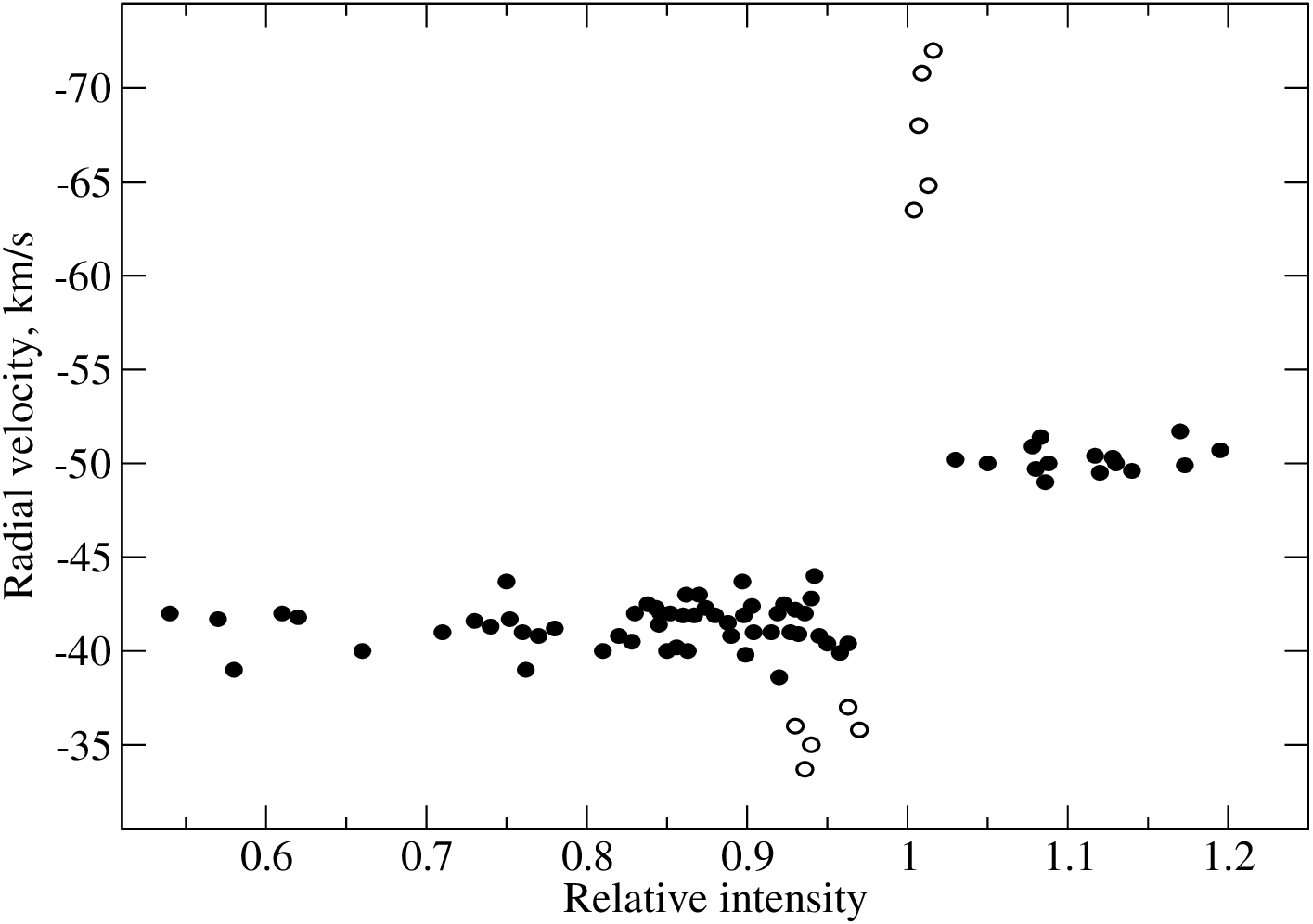}
\caption{Dependence of the radial velocity V$_r$ on residual intensity $r$ for the lines in the spectrum of
        IRAS\,01005 taken on May 29, 2013. Each symbol corresponds to an individual line. The filled circles 
        correspond to the forbidden emissions ($r > 1.0$, velocities measured by the peaks) and  He\,I and ion 
        absorptions ($r< 1.0$, velocities measured by the cores), and the open circles---to the emission 
        (V$_r< -60$~km\,s$^{-1}$) and absorption (V$_r>-40$\,km\,s$^{-1}$) components of  Fe\,III line profiles. }
\label{fig3}
\end{figure}

The accuracy of single-line velocity measurements is determined by the noise ($S/N$) and the gradient of residual 
intensity in the line profile ($\Delta r/\Delta\lambda$). As is evident from the example shown in Fig.\,\ref{fig2}, 
the profiles in the spectrum of IRAS\,01005 vary with time and from line to line. The accuracy of velocities based 
on individual-line measurements can be estimated from the scatter of the corresponding symbols in
Fig.\,\ref{fig3}, which shows the V$_r(r)$ dependences in the spectrum taken on May 29, 2013. The average 
velocity errors for this spectrum, which has the $S/N$ ratio typical of our data and the broadest wavelength coverage, 
are equal to 0.5\,km\,s$^{-1}$, about 1\,km\,s$^{-1}$, and about 3\,km\,s$^{-1}$ for the forbidden emissions, 
the absorption cores, and the weak emission components of  Fe\,III lines respectively.

\begin{figure}
\includegraphics[angle=0,width=0.35\columnwidth]{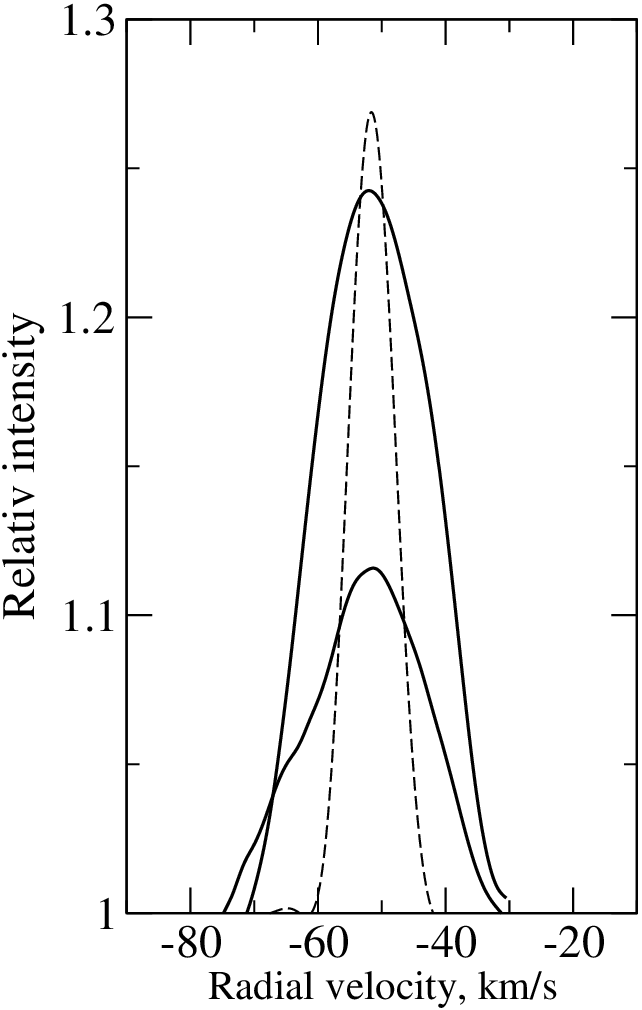}
\caption{Profiles of the  $[$N\,I$]$\,F1 5198\,\AA{} emissions and the weaker  O\,I\,(22) 6046\,\AA\ emission
         feature in the spectrum of IRAS\,01005 taken on August 21, 2013 compared to the profile of the 
         $[$O\,I$]$ F3 5577\,\AA{} telluric emission (shown by the dashed line shifted along the  V$_r$ axis).} 
\label{fig4}
\end{figure}

The least complex (symmetric) and stable profiles among the narrow emissions are those of the [N\,I], [N\,II], [O\,I], 
[S\,II], and [Fe\,II] forbidden lines. The presence of forbidden  [N\,II] and [S\,II] emissions is indicative of the 
onset of the ionization of the circumstellar envelope and of the closely imminent planetary nebula stage. 
The average radial velocities (measured on different dates from different sets of these lines) are listed in column~5
of Table\,\ref{spectra}. Temporal variations of the velocities of forbidden lines are close to those of interstellar
absorption features, i.e., to the measurement errors. The velocity appears to have remained constant during our observations.
Averaging over the entire dataset yields a value of $-50.5 \pm 0.2$~km\,s$^{-1}$. In Fig.\,\ref{fig4} we compare 
the profile of one of the stellar lines, [N\,I]~5198\,\AA{}, with that of the  [O\,I]~5577\,\AA\ telluric emission: 
the former is appreciably broader and has a half-width of about 10\,km\,s$^{-1}$.

\begin{figure}
 \vspace{1.3mm}
\includegraphics[angle=0,width=0.35\columnwidth]{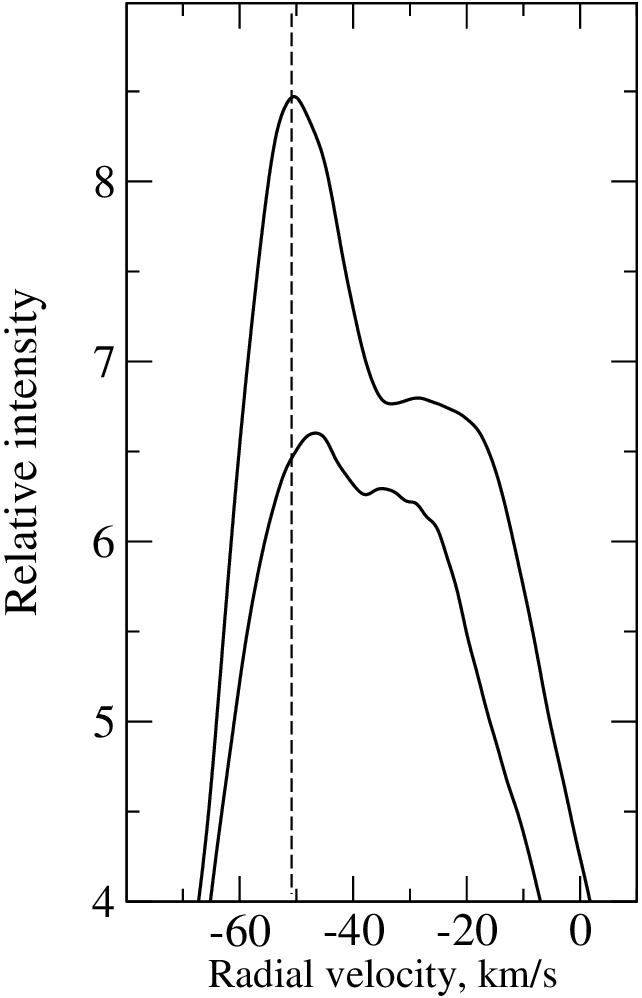}
\caption{The peaks of the H$\alpha$ profiles in the spectra of IRAS\,01005 (top to bottom): August 21,
         2013 and the average profile of November 22 and 25, 2002. The vertical dashed line indicates 
         the position of forbidden emissions.} 
\label{fig5}
\end{figure}

Unlike the radial velocities, the intensities of the forbidden lines exhibit small variations (with a maximum amplitude of
residual intensities equal to  5--6\%). The reality of these variations is supported by the fact that they occur
synchronously in different lines, by their absence over a two--day period, and the fact that relative intensity variations of the
lines in the [N\,I]~5198, 5200\,\AA, [Fe\,II]\,4814, 5158\,\AA\, and other doublets, which are limited only by observational errors,
are substantially smaller 2--3\%. The part of the spectrum of IRAS\,01005 available  to us contains several weak
O\,I emissions. Note that the very strong OI\,(1)\,7773\,\AA\ triplet is observed in absorption~[\cite{01005}]. Its
intensity might suggest that weaker lines --- members of multiplets (9)~(6454--56\,\AA\AA{}) and (10)~(6156--58\,\AA\AA{}) --- 
should  also have the form of absorptions with the depths of R$\approx 0.02$, however, they cannot be distinguished from noise in 
our spectra and are possibly filled with emissions. Pure emissions belong to the multiplets (5)~4368\,\AA, (23)~5958\,\AA\, 
and (22)~6046\,\AA{}, and their mean residual intensities are equal to  1.04, 1.05, and 1.11 respectively. We show the 
profile of the latter line in Fig.\,\ref{fig4}. The width and mean radial velocity ($-50.8 \pm 1$\,km\,s$^{-1}$) of the  
O\,I emission are close to the corresponding parameters of the forbidden lines. The fact that forbidden lines and  O\,I 
emissions form in the slowly expanding stellar envelope suggests that the latter should also be responsible for the 
formation of narrow emission components of the complex profiles of the first members of the Balmer series. During
our observations the wind components of these features showed appreciable intensity variations, but maintained a P\,Cyg\,III
type profile according to Beals with the red emission peak much higher than the blue one. Figure\,\ref{fig5} shows
examples of the top parts of complex H$\alpha$ profiles and the last column of Table\,\ref{spectra} lists the radial
velocities of the absorption and emission extrema of their wind components based on our entire observed data.

Visual inspection shows that variations of the envelope flux are more pronounced in hydrogen lines than in 
forbidden~O\,I emissions. Quantitative estimates are hindered by the blending of envelope and wind emissions, however, 
they can be performed by comparing all the available  H$\alpha$ profiles. The range of peak intensities of narrow 
components amounts to about  20\% if the intensity variations at the corresponding wavelengths in the wind
components are taken into account. Figure\,\ref{fig5} shows the upper parts of the H$\alpha$ profiles with relatively
strong (August 21, 2013) and weak  (November 22--25, 2002) envelope components. As is evident from these profiles, the
strongest (and hence least deformed by blending) narrow~H$\alpha$ emissions are more or less as wide as the forbidden 
emissions and are minimally shifted relative to them. As the component weakens, its peak shifts redward with its 
radial velocity changing from  $-50$\,km\,s$^{-1}$ on August 21, 2013 to $-43$\,km\,s$^{-1}$ on November 22--25, 2002.

The Si\,II lines also appear to form, at least partially, in the envelope. This is primarily evidenced by 
the fact that in the part of the spectrum of IRAS\,01005 available to us almost all these
lines appear in emission (the only exception is the 4128 and 4131\,\AA{} absorption--emission doublet). 
Furthermore, averaging over the entire radial velocity data for the emission peaks yields
a value close to  $-50$\,km\,s$^{-1}$. However, Si\,II lines differ from the envelope emissions considered 
above by the shape and variability of their profiles. The radial velocities of their peaks vary with time 
from $-46$\,km\,s$^{-1}$ to $-54$\,km\,s$^{-1}$, and their residual intensities and halfwidths---by 10\% 
and 30\% respectively. The time scale of these variations can hardly be determined from our data, however,
note that these variations do not exceed the measurement errors at least over two-day time intervals 
(November~25--27, 2002; November 13--15, 2005; and November~3--5, 2008).

\begin{figure}
 \vspace{1mm}
\includegraphics[angle=0,width=0.35\columnwidth]{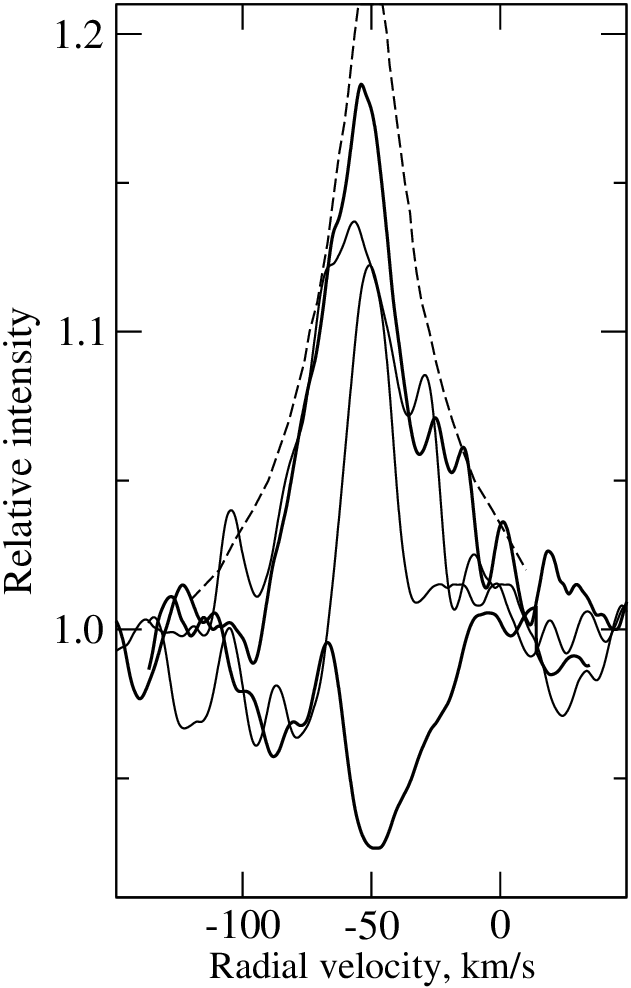}
\caption{Variations of  Si\,II profiles in the spectrum of IRAS\,01005. The bold lines show the  Si\,II\,(2)
         6347\,\AA\ emission and the average of the absorption--emission Si\,II\,(3) 4128 and 4130\,\AA{} 
         profiles observed on May 29, 2013. The thin lines show the Si\,II\,(2) 6347\,\AA{} profiles of
         November 15, 2005 (left) and April 13, 2003 (right). The dashed line shows the line enveloping of 
         all the Si\,II 5979\,\AA{} profiles.} 
\label{fig6}
\end{figure}

Figure\,\ref{fig6} shows examples of  Si\,II line profiles. The dashed line shows the line enveloping  of the 17
profiles of the strongest line, Si\,II~5979\,\AA{}, shifted along the V$_r$ axis. It resembles the silhouette of 
Mount Fuji on Japanese woodblock prints: a sharp peak ($r\approx 1.22$), incurved slopes ($\Delta$V$_r \approx \pm$20\,km\,s$^{-1}$
at half maximum), and a broad base (up to $\pm 90$\,km\,s$^{-1}$). The profile of the Si\,II~6347\,\AA\ line becomes sharply
asymmetric from time to time: one of its slopes ``sags,'' sometimes even below the continuum level (e.g., the blue slope on
April 13, 2003). These profile deformations may be due, as in the case of hydrogen lines, to  variations of their absorption
(photospheric) components and also to variations of the geometry and kinematics at the wind base. An inspection of columns 6 and 7
of Table\,\ref{spectra} shows that the profiles of April 13, 2003 and November 15, 2005 compared in Fig.\,\ref{fig6} correspond
to almost equal absorption-feature radial velocities.

\begin{figure}
\vspace{1mm}
\includegraphics[angle=0,width=0.35\columnwidth]{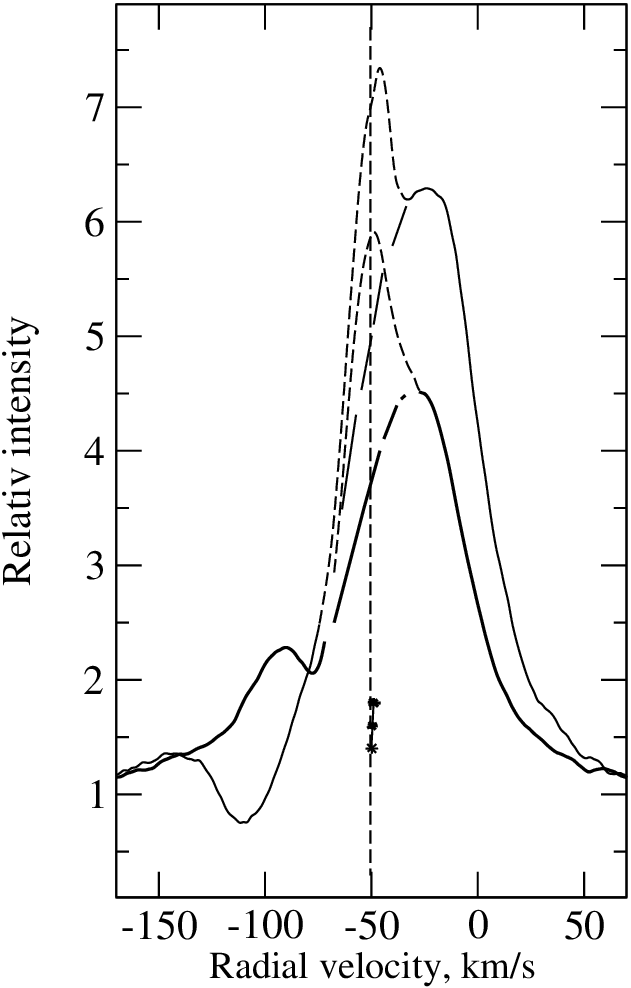}
\caption{H$\alpha$ profiles in the spectrum of IRAS\,01005 taken on March 9, 2004 (the thin line) and January
         18, 2005 (the bold line). The dashes show the envelope emissions, and the long-dashed lines show the 
         fragments of wind profiles under those emissions. The vertical dashed line marks the radial velocity 
         determined from forbidden emissions, and the chain of crosses next to it---the bisector of the lower 
         part of the profile of January 18, 2005.} 
\label{fig7}
\end{figure}

As we pointed out above, the H$\alpha$ lines, free from superimposing narrow envelope emissions, have P\,Cyg\,III-type,
i.e., almost wind-type, profiles. Figure\,\ref{fig7} illustrates this fact for two wind intensity levels (the residual
intensity of the red H$\alpha$ and H$\beta$ emission peaks varies by 50\% and 70\% over our entire observed data set). 
The main emission peaks and absorption depressions are located on opposite sides of the vertical line that marks the 
velocity determined from forbidden emissions ($-50.5$~km\,s$^{-1}$). As is evident from the last two columns of 
Table\,\ref{spectra}, for H$\alpha$ and H$\beta$ this also remains true for all our remaining spectra.
Note that both the shifts $\Delta$V$_r$\,=\,(V$_r$+50.5\,km\,s$^{-1}$) and their scatter are small: 
$-74 <\Delta$V$_r< -27$\,km\,s$^{-1}$ for the H$\alpha$ absorption minimum and 18$<\Delta$V$_r<$31\,km\,s$^{-1}$ 
for the main emission peak. Such a relatively low wind velocity, as well as the high intensity of the
red emission peak (it exceeds the continuum level by up to a factor of seven), are typical of hypergiants rather 
than classical supergiants. Hence IRAS\,01005 represents a case of spectral mimicry of a low  mass post--AGB star 
masquerading as a massive hypergiant.

The deviations of the wind structure from spherical symmetry are small. This is evidenced by the symmetry in the lower 
parts of the H$\alpha$ profiles (1.2$<r<$1.8), the fact that the radial velocities V$_r$ inferred from them are close 
to $-50$\,km\,s$^{-1}$, the inverse correlation between the intensities of the blue and red emission peaks, as well as the
inverse correlation between the depth of the absorption minimum and the height of the red emission peak and the direct 
correlation between the absorption depth and its blueshift. The conclusion about insignificant deviations of the wind 
from spherical symmetry is consistent with the shape of the circumstellar envelope of IRAS\,01005. 
Si\'odmiak et al.~[\cite{Siodmiak}] classified the high-resolution Hubble Space Telescope image of
this object as belonging to the SOLE morphological type, which is dominated by the flux of the central star. 
The circumstellar envelope is irregular and contains several lobes of different size.

\begin{figure}
 \vspace{1mm}
\includegraphics[angle=0,width=0.35\columnwidth]{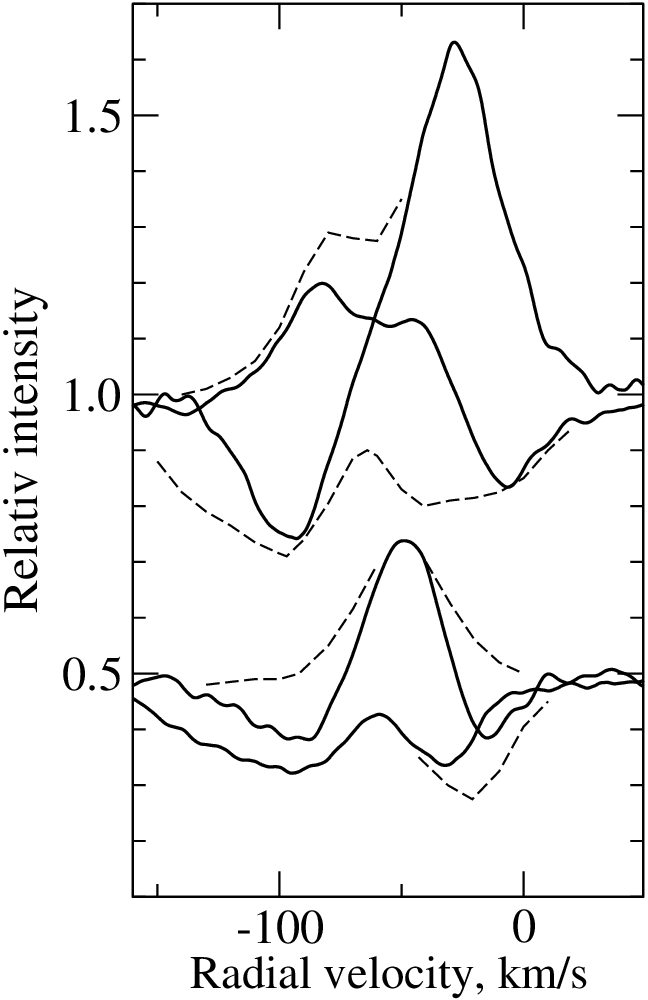}
\caption{Profiles of the  He\,I lines in the spectrum of IRAS\,01005: 5876\,\AA{} (top) and 5016\,\AA{}
        (shifted downwards by 0.5). The solid lines show the profiles with the highest and lowest emission 
        intensity (taken on April 13, 2003 and November 13, 2005 for 5876\,\AA{}, and on December 2, 2002 and
        May 29, 2013 for 5016\,\AA{}, respectively), and the dashed lines show the lines enveloping of the 
        corresponding sets of profiles.}
\label{fig8}
\end{figure}

The He\,I profiles are more diverse. Figure\,\ref{fig8} shows pairs of most dissimilar He\,I~5876 and 5016\,\AA{} line
profiles. The He\,I~5876\,\AA\ line, which is the strongest in the visible part of the spectrum, forms higher than 
the other He\,I lines, and therefore it is not surprising that in 12 out of our 20 spectra its profile reproduces 
the direct P\,Cyg profile of the H$\beta$ or H$\alpha$ line and follows its intensity variations. 
This similarity persisted, e.g., from November 22 through December 3, 2002. Another case is illustrated in
Fig.\,\ref{fig8} by the profile of April 13, 2003, where the shifts of the extrema (i.e., the $\Delta$V$_r$ quantity
introduced above) are equal to $-45$~and~$22$\,km\,s$^{-1}$, or, for H$\alpha$, $-50$ and $31$\,km\,s$^{-1}$. 
However, such similarity often disappears. In the profile of November 13, 2005, shown in the same figure, 
the main emission peak is on the blue rather than the red side of the absorption depression with the
shift equal to $\Delta$V$_r\approx -32$\,km\,s$^{-1}$ and $-6$\,km\,s$^{-1}$ for the former and latter respectively. 
The corresponding shifts for H$\beta$ are equal to $+17$ and $-25$\,km\,s$^{-1}$. Note that this is not the case 
of the direct P\,Cyg profile of the He\,I~5876\,\AA\ line becoming an inverse P\,Cyg profile, because both 
the emission and absorption features are located on the same side of the ``$-50$~km\,s$^{-1}$ line.''

The profile of the weaker  He\,I\,5016\,\AA{} line is less complex and more stable, it contains a single and 
rather narrow emission component (with an average halfwidth of 18\,km\,s$^{-1}$). The most variable of its 
parameters is the residual intensity, which, according to our data, varies by about 30\%, whereas the 
radial velocity oscillates near the ``$-50$~km\,s$^{-1}$ line'' from $-62$ to $-43$\,km\,s$^{-1}$.

\begin{figure}
 \vspace{1mm}
\includegraphics[angle=0,width=0.35\columnwidth]{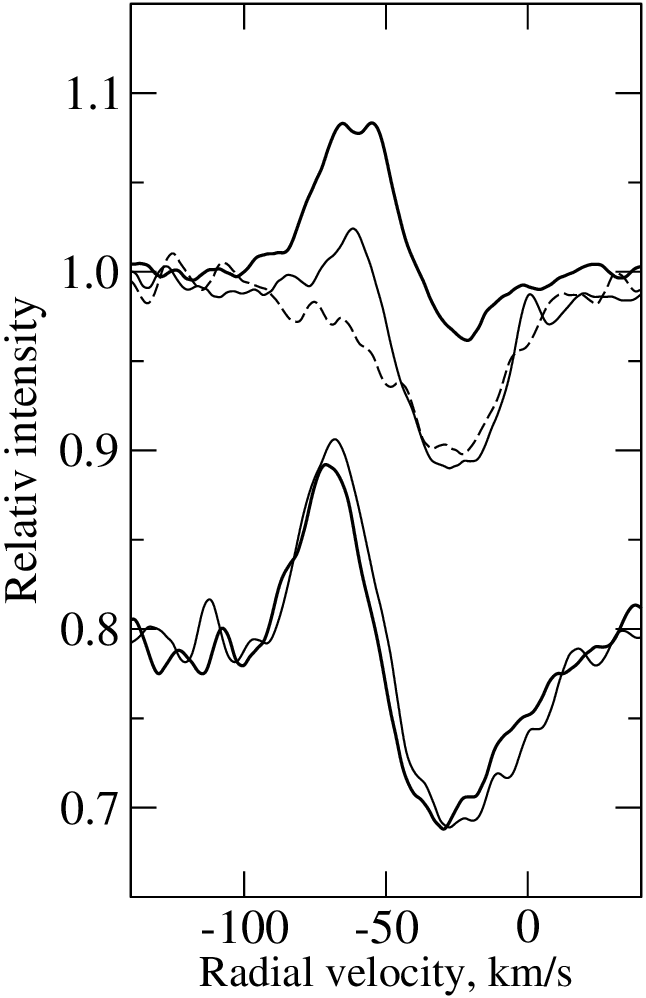}
\caption{The similarity of emission-absorption profiles in the spectrum of IRAS\,01005 of August 21, 2013. 
         Top:  Fe\,III (the average for the 5127 and 5156\,\AA{} features shown by the bold line) and He\,I 
         (5048\,\AA{}, the thin line) profiles; for comparison we also show the profile of the O\,II 4267\,\AA{} 
         absorption (the dashed line). Bottom: the C\,II 4267\,\AA{} (the bold line) and He\,I 4713\,\AA{} 
         (the thin line) profiles, shifted downward by 0.2 along the vertical axis.}
\label{fig9}
\end{figure}

The profiles of even weaker  He\,I lines become increasingly similar to the emission-absorption profiles
of the heavier elements: C\,II, N\,II, and Fe\,III (Fig.\,\ref{fig9}). The latter features have the most 
typical profiles and we therefore list the velocities of their extrema in column~8 of Table\,\ref{spectra} 
(the similarity of the profiles allowed us to replace the unavailable reliable Fe\,III lines by He\,I line 
measurements in two cases). Interpreting the profiles of the He\,I~4713\,\AA\ and other lines as inverse
P\,Cyg profiles, i.e., as a manifestation of the global contraction of their formation layers, appears 
to be justified in some cases, however, most of our spectra show only an emission superimposed onto the 
blue slope of the absorption profile. The upper part of Fig.\,\ref{fig9} demonstrates this fact by an example 
of the profile of the pure absorption  O\,II~4676\,\AA\ line of similar intensity.

\begin{figure*}
 \vspace{5mm}
\includegraphics[angle=0,width=0.65\textwidth, bb=5 168 633 520,clip]{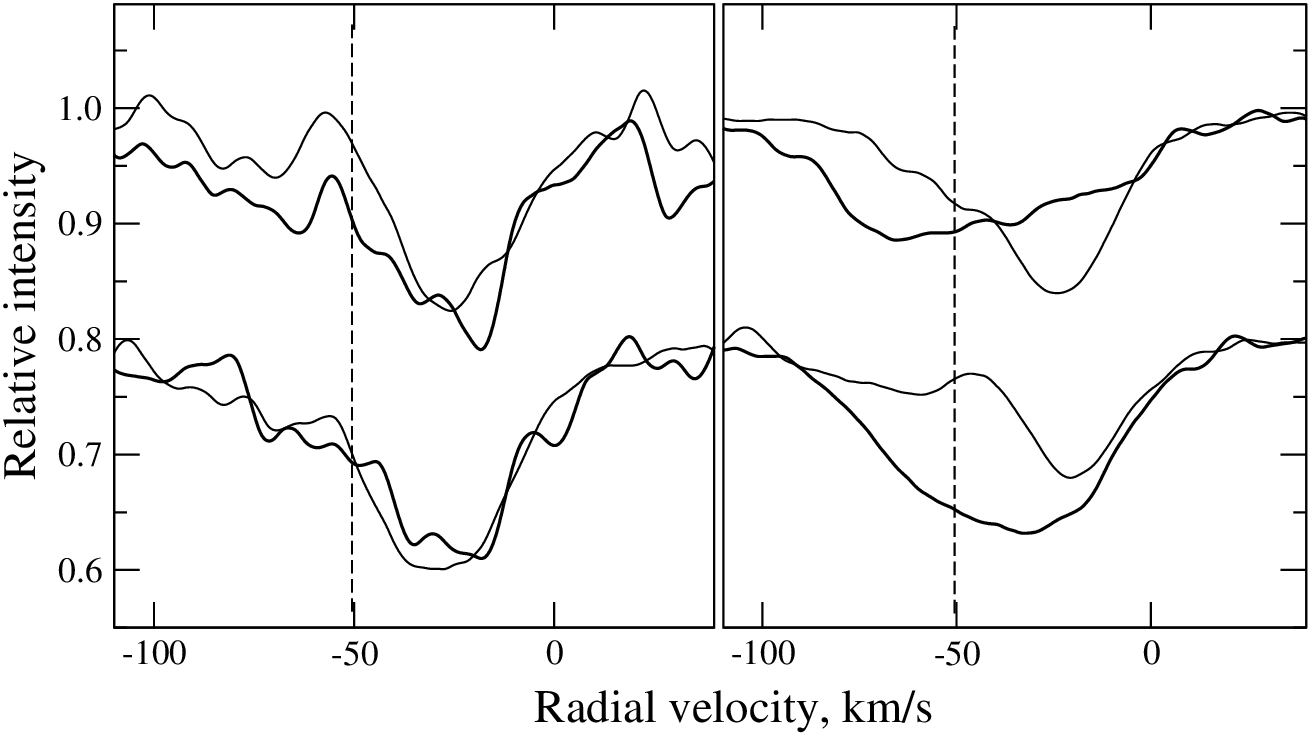}
\caption{Temporal variations of the He\,I 5048\,\AA{} (top) and N\,II 5679\,\AA{} (shifted downward by 0.2)
         profiles in the spectrum of IRAS\,01005 over a six-hour (December 2/3, 2002, the left-hand side of 
         the figure) and two-day (November 3 and 5, 2008, the right-hand side of the figure) time periods. The 
         thin and bold lines show the earlier and later profile shapes respectively.} 
\label{fig10}
\end{figure*}

The lines that form at the wind base and in the transition layer between the wind and the photosphere are also 
variable in all their parameters. Our data demonstrates conclusively the profile variations at least over a 
two--day time interval. This is immediately apparent from Fig.\,\ref{fig10}. Whereas the differences between 
the  He\,I 5048\,\AA\ and N\,II 5679\,\AA\ line profiles appearing over a six-hour time interval (the
left-hand fragment of the figure) are still close to the measurement errors, the corresponding variations over 
a two--day time interval (the right-hand fragment) become significant and quite measurable: thus the equivalent 
width of the  N\,II~5679\,\AA\ line increased from  0.15 to 0.22\,\AA{}.

More or less pure photospheric absorptions are usually asymmetric: their cores are redshifted relative to the wings.
We therefore measured the radial velocities of these lines not only by their cores but also by the wings of their 
rofiles, and list the corresponding values in columns~7~and~6 of Table\,\ref{spectra} respectively. Each value is a
result of averaging over many absorptions of the spectrum based on the corresponding plot of the type shown in
Fig.\,\ref{fig3}. Note that the  V$_r(r)$ dependence zone may have a nonzero slope and therefore the adopted 
result is equal to the limit of  V$_r$ as residual intensity tends to $r$\,=\,1.0.

\begin{figure*}
\includegraphics[angle=0,width=0.75\textwidth, bb=0 0 705 525, clip]{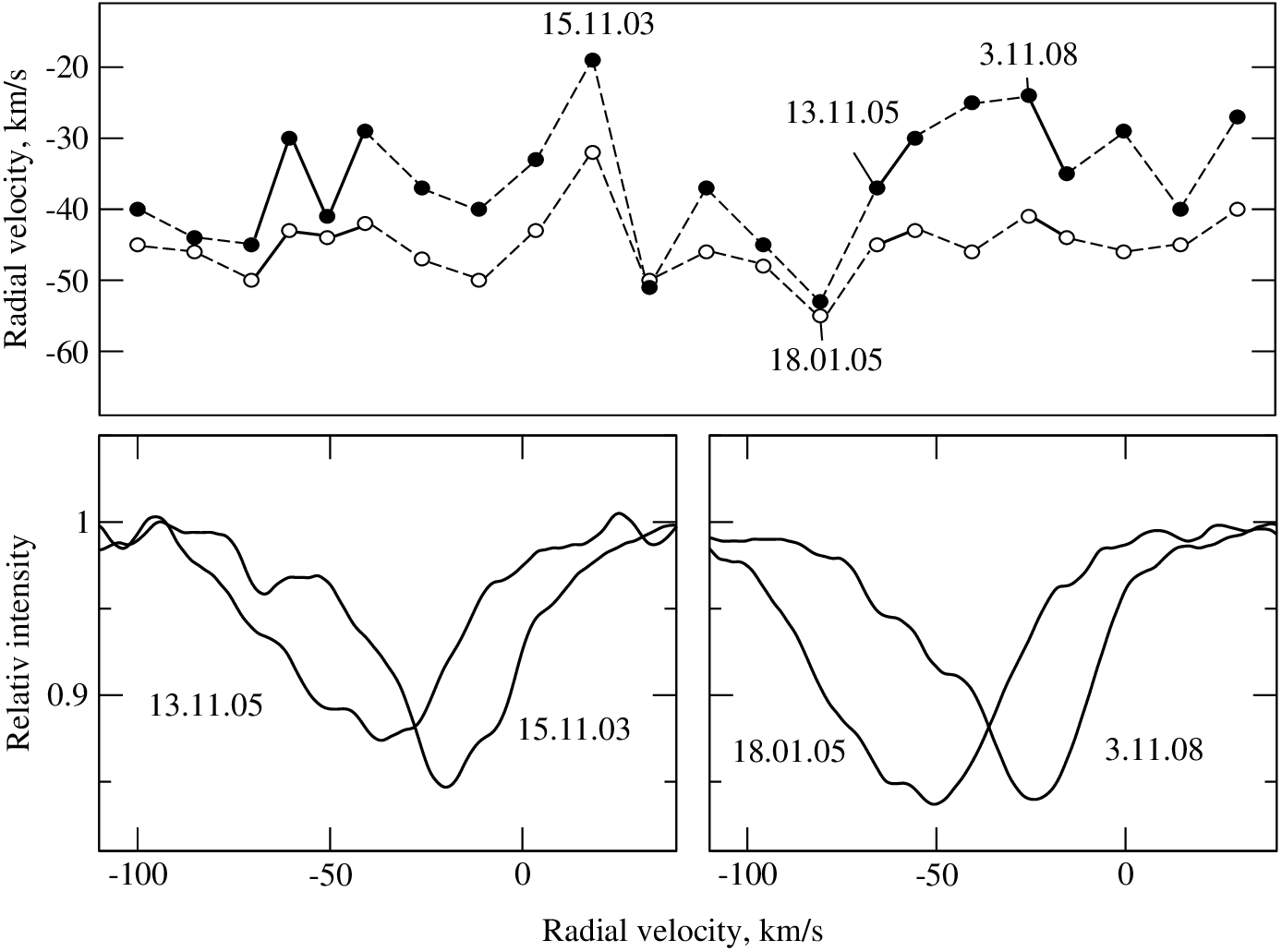}
\caption{Top:  variation of the radial velocities of weak absorptions in the spectrum of IRAS\,01005
         measured by their cores (the filled circles) and wings (the open circles) from day to day 
         (between January 25, 2002 and August 21, 2013). The data for close dates are connected by 
         solid lines; in other cases---by dashed lines. The dates are marked for which the line profiles 
         are given in the bottom panel. Bottom: a comparison of the N\,II\,4630\,\AA{} (left) and N\,II\,5679\,\AA{} (right)
         profiles for dates with large discrepancies between the measured radial velocity values. } 
\label{fig11}
\end{figure*}

Figure\,\ref{fig11} shows graphically the radial velocities from columns~6 and~7 of Table\,\ref{spectra}. Their temporal 
variations should be viewed relative to the radial velocity of the star as a whole, i.e., the systemic radial 
velocity V$_{\rm sys}$. It is reasonable to set this velocity equal to the radial velocity of forbidden emissions, 
i.e., adopt V$_{\rm sys}=-50.5$\,km\,s$^{-1}$. This assumption is supported by the fact that the radial velocity
values V$_r$ for the wings, which correspond to the deepest accessible atmospheric layers, are closer to the adopted 
V$_{\rm sys}$ than the  V$_r$ values for the line cores. More importantly, absorptions with V$_r\approx$V$_{\rm sys}$ 
are least deformed by emission features or are even totally free of them (see the lower panels in Fig.\,\ref{fig11}, 
where we compare the profiles of two  N\,II lines at the epochs of large discrepancies between the inferred radial 
velocity values). According to our data, the spread of V$_r$  estimates for the line cores
amounts to 34\,km\,s$^{-1}$, and can partly be due to the deformations of the profiles by variable emission features. 
The spread of  V$_r$  estimates for the wings is smaller, 23\,km\,s$^{-1}$, and can be interpreted as a result 
of pulsations and/or hidden binarity of the star even despite the low accuracy of the corresponding measurements. 
The hypothesis that the star may be pulsationally unstable is consistent with a similar interpretation of the 
rapid variability of the photometric parameters of the star~[\cite{Arkhip2013}].

\subsection{Binarity and pulsations of post-AGB stars}

Many of the PPN candidates exhibit light and radial velocity variations on time scales of several hundred days 
which may be indicative of their binary nature. Conclusive evidence for orbital motions has been obtained 
for several optically bright stars undergoing the PPN stage. For example, the high-latitude supergiants 
89\,Her~[\cite{Ferro,Waters}], HR\,4049, HD\,44179, and HD\,52961~[\cite{Winck95}] are shown to be binary, 
and the above authors determined the orbital elements and proposed the corresponding binary models for these
systems. The nature of the companions in suspected binary post--AGB stars so far remains unknown because of the lack
of their direct manifestations in the continuum or spectral lines: all the known post--AGB binaries are of 
the SB1 type. A possible companion can be either a very hot object or a low-luminosity main-sequence star. 
It can also be a white dwarf, like in the case of Ba--stars.

According to theoretical computations of Gautschy~[\cite{Gautschy}], pulsations are a typical
feature of  post-AGB stars throughout a wide temperature interval, $3.8 \le \log T_{\rm eff} \le 4.9$. 
Manifestations of atmospheric pulsations were earlier found for a number of stars at the PPN
stage. Such is the case of the well-studied semiregular variable CY\,CMi (the optical component of the IR--source 
IRAS\,07134+1005). The radial velocity variability of this star was first suspected in~[\cite{Kloch95}] from 
a comparison of the published data with those obtained with the 6--m telescope. Somewhat later Lebre et al.~[\cite{Lebre}] 
used the Fourier technique to analyze the variability of the H$\alpha$ profile in the spectrum of  CY\,CMi  
and the corresponding radial velocity data, and concluded that the atmosphere of this star has complex
pulsation driven dynamics. Barth\`es et al.~[\cite{Bart}] acquired an extensive set of quality spectra of 
CY\,CMi spanning an eight-year period and concluded that the radial velocity of the star varies with a 
half-amplitude and main period of  2.7\,km\,s$^{-1}$ and P\,=\,36$\fd8\pm 0.2$ respectively. Photometric 
variations have the same period and a very small amplitude of $0\fm02$.

Klochkova et al.~[\cite{56126-atlas,56126-AZh}] analyzed the kinematics of the atmosphere and envelope of 
CY\,CMi by studying spectral features of various intensity, found the kinematic pattern to be variable, 
and pointed out that the variations of the radial velocities measured from extremely weak  absorptions 
possibly indicate that the star is a binary, but may also be a manifestation of low-amplitude pulsations 
in circumphotospheric layers. To determine conclusively whether CY\,CMi is a binary it is important 
to monitor the V$_r$ variations over several years acquiring one to two spectra every month.

Hrivnak et al.~[\cite{Hrivnak2013}] analyzed the pulsations in the atmosphere of V448\,Lac and in that of a related
post--AGB star V354\,Lac (IRAS\,22272+5435) based on long-term observations of light, color, and radial-velocity 
variations. The above authors determined the periods, period ratios, and pulsation amplitudes for these stars and found that
the pulsation properties of post--AGB stars are not always consistent with the results of theoretical computations for this
evolutionary stage~[\cite{Fokin}] and differ from the corresponding properties of classical Cepheids. The stars CY\,CMi,
V354\,Lac, and V448\,Lac, which were found to be pulsationally unstable, are a part of the group of F--type supergiants
enriched in carbon and s--process heavy metals synthesized during prior evolution. The basic information about 
this group of related PPNe can be found in~[\cite{Hrivnak2010,Envelopes}]. Klochkova et al.~[\cite{05040}] 
found evidence for variations of the velocities of the absorption features in the spectrum of a hotter 
post--AGB star --- the A--type supergiant ${\rm BD}\!+\!48\degr 1220={\rm IRAS}\,05040\!+\!4820$.

\subsection{Lines of the Na\,I doublet and DIBs}

The heliocentric radial velocities of the five main components of the Na\,I D lines (Fig.\,\ref{fig1}) averaged over 
our entire dataset are equal to $-72.5$, $-65.3$, $-52.2$, $-27.7$, and $-10.2$\,km\,s$^{-1}$, and hence agree 
within the quoted errors with the results reported in our previous paper~[\cite{01005}] and with those reported 
in a recent paper~[\cite{Iglesias}]. The weak component of the Na\,I D lines 
whose position corresponds to V$_r=-52.2$\,km\,s$^{-1}$ forms in the stellar atmosphere, whereas the two
red components are of interstellar origin and form in the Local arm of the Galaxy. The blue component
(V$_r=-65.3$\,km\,s$^{-1}$) may also form in the interstellar medium, in the Perseus arm. This hypothesis is 
supported by the presence of a similar interstellar component with V$_r \approx -63$\,km\,s$^{-1}$ in 
the spectra of the B-type stars  HD\,4841, HD\,4694, and Hiltner~62~[\cite{00470}], which are members of the 
Cas\,OB7 association. The Galactic latitudes of these stars are close to that of IRAS\,01005. This hypothesis
leads us to conclude that IRAS\,01005 should be a very distant object. The distance to the Cas\,OB7 association 
is equal to $d=2.5$\,kpc~[\cite{Cazzolato}] and can be viewed as a lower estimate for the distance to IRAS\,01005.

Note that Klochkova et al.~[\cite{23304new}[ also found a strong interstellar Na\,I\,D line component with a close velocity
V$_r=-61.6$\,km\,s$^{-1}$ in the spectrum of the faint G-type supergiant identified with the IR--source IRAS\,23304+6347, 
located close to the Galactic plane at a Galactic longitude that differs by about~$10\degr$ from that of
IRAS\,01005.

As for the bluemost component of the Na\,I D lines with the velocity of V$_r=-72.5$, it seems quite logical 
to conclude that it forms in the expanding circumstellar envelope in the system of the IRAS\,01005 source. 
In this case we derive an expansion velocity of V$_{\rm exp}\approx 22$\,km\,s$^{-1}$, which is
typical for PPNe (see~[\cite{Loup,Envelopes}]).

The spectra of the B--type stars  HD\,4841, HD\,4694, and Hiltner\,62 contain DIBs with velocities ranging from
$-11$~to~$-14$\,km\,s$^{-1}$~[\cite{00470}]. We also identified four DIBs in the spectra of  IRAS\,01005 at
5780, 5797, 6195, and 6613\,\AA{}. The positions of these features correspond to a radial velocity of 
V$_r=-13.5$\,km\,s$^{-1}$, which agrees with the velocity inferred from three bands by Iglesias--Groth and
Esposito~[\cite{Iglesias}] and with the measurements made by Luna et al.~[\cite{Luna}] from the 5780 and
6613\,\AA{} bands. The measurements of other DIBs in IRAS\,01005 spectrum reported by Luna et al.~[\cite{Luna}] have large errors. 
The same is true for the measurements of the positions of the components of the D lines of the  Na\,I doublet 
by the same authors.

\subsection{Spectral type of the IRAS\,01005 central star}

We derive a spectral type of B1.5$\pm 0.3$ for the star based only on the weak and minimally deformed  C\,II/III,
N\,II/III,  O\,II/III,  Si\,III/IV, and other absorptions, and a direct comparison of the spectrum of IRAS\,01005 
with those of V1853\,Cyg (B1\,Iae), 9\,Cep (B2.3\,Ib), and other early type supergiants from the 
atlas~[\cite{Sarkisyan}]. The H\,I and He\,I lines yield a luminosity class Ib for IRAS\,01005. Klochkova et
al.~[\cite{01005}] found for IRAS\,01005 a similar spectral--type estimate ${\rm B}1.7 \pm 0.5$. Note that the
spectral types of metal--poor stars with the chemical composition altered during evolution determined by comparing 
the spectra of these objects with those of Population-I stars may be fraught with systematic errors.

The hot post--AGB supergiants V1853\,Cyg~[\cite{V1853Cyg}], V886\,Her~[\cite{V886Her}], LS\,III~$52\degr24$~[\cite{Sarkar2012}], 
and several more stars in the southern sky are related objects for IRAS\,01005~[\cite{Sarkar2005,Mello2012}].
Gauba and Parthasarathy~[\cite{Gauba2004}] report a list of 15 hot post--AGB stars, analyze the specific features 
of their IR--spectra, and determine the parameters of their circumstellar envelopes. Of special interest among the hot
post-AGB stars is the high-latitude Be--star V886\,Her (IRAS\,18062+2410), which is similar to IRAS\,01005 
in terms of basic parameters. An analysis of extensive observational data for V886\,Her suggests that it is 
rapidly evolving toward the PN stage~[\cite{Arkhip1999,Partha2000,Arkhip2007}].

\section{Conclusions}\label{conclus}

Based on 23 high-resolution spectra (R\,=\,60\,000) of the IR--source IRAS\,01005+7910 that we took with the 
6--m telescope of the Special Astrophysical Observatory, we determined  the spectral type and luminosity class 
(B1.5$\pm 0.3$ \,Ib) of the central star, identified numerous spectral features, analyzed the variations of
their profiles and radial velocity.

We determined the systemic radial velocity of the star V$_{\rm sys}$\,=$-50.5$\,km\,s$^{-1}$ from the positions 
of symmetric and stable profiles of forbidden emission features of  [N\,I], [N\,II], [O\,I], [S\,II], and [Fe\,II]. 
The presence of forbidden emissions of [N\,II] and [S\,II] is indicative of the onset of the ionization of the 
circumstellar envelope and the approaching planetary nebula stage.

The span of radial-velocity estimates V$_r$ based on the line cores, which amounts to 34\,km\,s$^{-1}$, 
is partially due to the profile deformations caused by variable emissions. The span of V$_r$ estimates 
based on the line wings is smaller, about 23\,km\,s$^{-1}$, and can be due to pulsations and/or hidden binarity 
of the star. The deformations of the profiles of absorption-emission lines may be due to the variations of 
their absorption (photospheric) components with varying geometry and kinematics in the wind base. Our observed data
leads us to conclude that variations become quite measurable over a two--day time interval.

The H$\alpha$ lines have  P\,Cyg\,III-type wind profiles. We show that deviations of the wind from spherical symmetry 
are small. We find the wind velocity to be small (27--74\,km\,s$^{-1}$ at different observing epochs) and the 
red emission to have high intensity (exceeding the continuum level by up to a factor of seven) which is typical 
of hypergiants rather than classical supergiants.

According to our measurements, the heliocentric radial velocities of the five main components of Na\,I D lines are equal to
V$_r$\,=$-72.5$, $-65.3$, $-52.2$, $-27.7$, and  $-10.2$\,km\,s$^{-1}$ (Fig.\,\ref{fig1}) and agree with
the published data within the errors. The weak component at V$_r$\,=$-52.2$\,km\,s$^{-1}$ forms in the stellar 
atmosphere and the two redder components are of interstellar origin and form in the Local arm. The presence of the 
V$_r$\,=$-65.3$\,km\,s$^{-1}$ component, which appears to form in the interstellar medium of the Perseus arm, 
allows us to adopt d\,=\,2.5\,kpc as a lower estimate for the distance to IRAS\,01005. The bluemost component,
V$_r$\,=$-72.5$\,km\,s$^{-1}$, may form in the circumstellar envelope expanding at a velocity of 
V$_{\rm exp}\approx$22\,km\,s$^{-1}$,  which is typical of PPNe.

\subsection*{Acknowledgments}
This work was supported by the Russian Foundation for Basic Research (project No.~14--02--00291\,a).  The observations 
with the  6--m telescope were carried out with the financial support of the Ministry of Education and Science of the 
Russian Federation (contracts No.~16.518.11.7073 and 14.518.11.7070). 
This research has made use of the SIMBAD database, operated at CDS, Strasbourg, France, and  SAO/NASA Astrophysics Data System.

\clearpage

\begin{longtable}{l @{\quad}c @{\qquad}l@{\qquad}l}
\caption{Identification of the lines in the spectrum of IRAS\,01005, their residual intensities ($r$) and heliocentric
         radial velocities (V$_r$). The letter ``e'' indicates the emissions and emission components. Spectral variability 
         prevents averaging over several spectra, and therefore the 3530--4575\,\AA{}, 4590--5455\,\AA{}, and 5495--6731\,\AA{}
         wavelength intervals are represented by the spectra of November~15, 2003, November~13, 2005, and January~18, 2005,
         respectively.  The less reliable measurements are marked by colons.}      
\endfirsthead
\hline
\multicolumn{4}{l}{Tabl.\,2, continuation} \\ \hline
\endhead
\hline
\multicolumn{4}{r}{Tabl.\,2, to be continued} \\  \hline
\endfoot
\hline
\endlastfoot
\hline
Line        &  $\lambda$,     &     r   &   Vr,  \\ 
(multiplet) &    \AA{}       &         &   km\,s$^{-1}$ \\
\hline
 \multicolumn{4}{c}{\underline{15.11.03}} \\
He\,I (36)   &      3530.49   &       0.71:  &   $-22:$ \\
He\,I (34)   &      3554.43   &       0.68:  &   $-28:$ \\
He\,I (31)   &      3587.29   &       0.76   &   $-27 $ \\
He\,I (6)    &      3613.64   &       0.77   &   $-25:$ \\
He\,I (28)   &      3634.25   &       0.74   &   $-23 $ \\
H\,24        &      3671.48   &       0.94:  &   $-12:$ \\
H\,23        &      3673.76   &       0.88:  &   $-13:$ \\
H\,22        &      3676.36   &       0.87:  &   $-19:$ \\
H\,21        &      3679.35   &       0.83:  &   $-15:$ \\
H\,20        &      3682.81   &       0.81:  &   $-21:$ \\
H\,19        &      3686.83   &       0.79:  &   $-19 $ \\
H\,18        &      3691.56   &       0.73   &   $-16 $ \\
H\,17        &      3697.15   &       0.72   &   $-21 $ \\
H 16         &      3703.86   &       0.64:  &   $-15:$ \\
He\,I (25)   &      3705.02   &       0.61   &   $-26 $ \\
H\,15        &      3711.97   &       0.64   &   $-15 $ \\
H\,14        &      3721.94   &       0.64   &   $-13 $ \\
O\,II (3)    &      3727.33   &       0.87:  &   $-27:$ \\
He\,I (24)   &      3732.87   &       0.81   &   $-18:$ \\
H\,13        &      3734.37   &       0.65   &   $-12 $ \\
H\,12        &      3750.15   &       0.63   &   $-14 $ \\
He\,I (65)   &      3768.78   &              &          \\
H\,11        &      3770.63   &       0.65   &   $-18 $ \\
             &                &      e 0.72: &   $-43:$ \\
He\,I (64)   &       3784.86  &       0.92   &   $-23 $ \\
Si\,III (5)  &       3791.41  &       0.93   &   $-16:$ \\
H\,10        &       3797.90  &       0.64   &   $-26 $ \\
             &                &      e 0.69: &   $-40:$ \\
He\,I (63)   &       3805.74  &       0.90   &   $-23:$ \\
Si\,III (5)  &       3806.54  &       0.90:  &   $-25:$ \\  
He\,I (22)   &       3819.64  &     e 1.00   &   $-70 $ \\
             &                &       0.61   &   $-26 $ \\
He\,I (62)   &       3833.55  &       0.83   &   $-20 $ \\
H\,9         &       3835.38  &       0.68   &   $-20:$ \\
             &                &     e 0.80:  &   $-32:$ \\
He\,I (61)   &       3838.10  &              &          \\ 
N\,II (30)   &       3838.37  &              &         \\ 
N\,II (30)   &       3842.18  &       0.96:  &   $-23:$ \\
N\,II (30)   &       3847.41  &       0.98:  &   $-33:$ \\
Si\,II (1)   &       3856.02  &     e 1.13   &   $-52 $ \\
Si\,II (1)   &       3862.60  &     e 1.10   &   $-48 $ \\
He\,I (20)   &       3867.50  &       0.84   &   $-23 $ \\
He\,I (60)   &       3871.79  &       0.81   &   $-24 $ \\
C\,II (33)   &       3876.2:  &       0.90   &         \\
O\,II (12)   &       3882.20  &       0.94:  &   $-24:$ \\
He\,I (2)    &       3888.65  &     e 1.01   &   $-77 $ \\
             &                &       0.66   &   $-31 $ \\
H\,8         &       3889.05  &       0.70   &         \\
O\,II (17)   &       3911.96  &       0.94:  &   $-28:$ \\
C\,II (4)    &       3918.98  &       0.84   &   $-24 $ \\
O\,II (17)   &      3919.29   &              &          \\
C\,II (4)    &      3920.69   &      0.76    &   $-26 $ \\
He\,I (58)   &      3926.53   &     0.79     &   $-23 $  \\
Ca\,II (1)   &      3933.66   &     0.48     &   $-70.5$\\
             &                &     0.30     &   $-12.4$ \\
He\,I (57)   &      3935.91   &     0.95     &   $-19:$  \\
O\,II (6)    &      3945.04   &     0.95     &   $-30:$  \\
O\,II (6)    &      3954.36   &     0.92     &   $-25:$  \\
N\,II (6)    &      3955.85   &     0.93:    &   $-29:$  \\
He\,I (5)    &      3964.73   &     0.70     &   $-24 $  \\
Ca\,II (1)   &      3968.47   &      0.61:   &   $-71.0$ \\
             &                &      0.45:   &   $-12.6$ \\
H$\epsilon$  &      3970.07   &      0.70:   &   $ -68:$ \\
             &                &    e 0.83:   &   $ -30 $  \\
O\,II (6)    &      3973.26   &        -     &   $ -21$ \\
O\,II (6     &      3982.71   &      0.92:   &   $ -30:$ \\
N\,II (12)   &      3995.00   &      0.80    &   $ -23$  \\
He\,I (55)   &      4009.27   &      0.73    &   $ -25$  \\
Fe\,III (45) &      4022.35   &      0.98:   &   $ -23:$  \\
He\,I (54)   &      4023.97   &      0.96    &   $ -25:$ \\
Fe\,III (53) &      4025.07   &      0.93    &   $ -21$  \\ 
He\,I (18)   &      4026.23   &    e 0.98    &   $ -73$  \\
             &                &       0.62   &   $ -27$ \\
N\,II (39)   &      4035.08   &      0.97    &   $ -33$  \\
N\,II (39)   &      4041.31   &      0.94    &   $ -31:$ \\
O\,II (10)   &      4069.8:   &       0.89   &          \\
O\,II (10)   &      4072.16   &      0.90    &   $ -28:$ \\
C\,II (36)   &      4074.6:   &       0.95:  &           \\
C\,II (36)   &      4075.85   &      0.86    &   $ -21$  \\
O\,II (10)   &      4075.87   &              &          \\
O\,II (10)   &      4078.86   &     0.96     &   $-27:$  \\
O\,II (10)   &      4083.91   &              &          \\
O\,II (10)   &      4085.11   &     0.95:    &   $-30:$  \\
Si\,IV (1)   &      4088.85   &      0.91    &   $-19:$  \\
O\,II (48)   &      4089.29   &              &          \\
O\,II (10)   &      4092.94   &      0.95    &   $-29:$  \\
H$\delta$    &      4101.74   &       0.78   &   $-70 $  \\
             &                &     e 0.98   &   $-34 $  \\ 
O\,II (20)   &      4103.02   &              &          \\
O\,II (20)   &      4104.9:   &              &          \\           
O\,II (20)   &      4110.50   &        0.96: &          \\
O\,II (21)   &      4112.03   &        0.97  &    $-30:$ \\
Si\,IV (1)   &      4116.10   &        0.96  &    $-27:$ \\
O\,II (20)   &      4119.22   &        0.91  &    $-26:$ \\
He\,I (16)   &      4120.82   &        0.77  &    $-21 $ \\
Si\,II (3)   &      4128.07   &     e 1.01:  &    $-67:$ \\
             &                &       0.88   &    $-29 $ \\
Si\,II (3)   &      4130.89   &      e 1.00: &    $-67:$ \\
             &                &        0.91  &    $-28 $ \\
O\,II (19)   &      4132.80   &         0.96:&    $-31:$ \\
He\,I (53)   &      4143.76   &      e 1.00  &    $-70 $ \\
             &                &        0.66  &    $-25 $ \\
O\,II (19)   &      4153.30   &         0.89 &    $-28 $ \\
He\,I (52)   &      4168.97   &        0.91  &    $-23 $ \\
N\,II (42)   &      4176.16   &        0.97  &    $-23:$ \\
O\,II (36)   &      4185.46   &      0.96    &    $-24:$ \\
O\,II (36)   &      4189.80   &      0.92    &    $-29 $ \\
N\,II (33)   &      4227.74   &      0.96:   &            \\
CH$^+$       &      4232.55   &      0.96    &    $-12.8$ \\
N\,II (48)   &      4241.78   &      0.95    &    $-26 $ \\
$[$Fe\,II$]$ F21&   4243.98   &   e 1.05     &    $-50 $ \\
O\,II (101)  &      4253.8:   &       0.91   &    $    $ \\
C\,II (6)    &      4267.14   &   e 1.05     &    $-68 $ \\
             &                &     0.77     &    $-18 $ \\
O\,II (67)   &      4275.52   &      0.93    &    $-28:$ \\
$[$Fe\,II$]$ F21&   4276.83   &   e 1.03:    &    $-51 $ \\        
S\,III (4)   &      4284.89   &      0.97:   &    $-19:$ \\
$[$Fe\,II$]$ F7 &   4287.40   &   e 1.15     &    $-50 $ \\
S\,II (49)   &      4294.40   &      0.97:   &    $-20 $ \\
CH           &      4300.32   &      0.96    &    $-12 $ \\
O\,II (54)   &      4303.83   &      0.95:   &           \\
C\,II (28)   &      4313.10   &      0.97:   &    $-15:$ \\
O\,II (78)   &      4313.43   &              &           \\
O\,II (2)    &      4317.14   &      0.88    &    $-22 $ \\
C\,II (28)   &      4317.26   &              &           \\
O\,II (2)    &      4319.63   &      0.83    &    $-21 $ \\
O\,II (2)    &      4325.76   &      0.97:   &           \\
C\,II (28)   &      4325.9:   &              &           \\
S\,III (4)   &      4332.71   &      0.97    &    $-27:$ \\
O\,II (2)    &      4336.86   &      0.93:   &    $-26 $ \\
H$\gamma$    &      4340.47   &   e 1.00     &    $-100$ \\
             &                &     0.84     &    $-72 $ \\
             &                &   e 1.32     &    $-30 $ \\
O\,II (2)    &      4345.56   &      0.86    &    $-20 $ \\
O\,II (16)   &      4347.42   &      0.95:   &    $-27:$ \\
O\,II (2)    &      4349.43   &     0.79     &    $-19 $ \\
O\,II (2)    &      4351.26   &     0.92:    &    $-30:$ \\
$[$Fe\,II$]$ F7&    4359.74   &  e 1.10      &    $-51 $ \\
S\,III (4)   &      4361.48   &     0.97:    &    $-26:$ \\
O\,II (2)    &      4366.89   &     0.89     &    $-23 $ \\
O\,I (5)     &      4368.25   &  e 1.04      &    $-51:$ \\
Fe\,III (122)&      4372.31   &     0.97:    &           \\
C\,II (45)   &      4372.4:   &              &           \\
C\,II (45)   &      4374.27   &     0.96:    &    $-27:$ \\
Fe\,III (4)  &      4382.51   &     0.97:    &    $-20:$ \\
He\,I (51)   &      4387.93   &  e 1.00:     &    $-85:$ \\
             &                &     0.61     &    $-25 $ \\
Fe\,III (4)  &      4395.76   &  e 1.03      &    $-62:$ \\
             &                &    0.94      &    $-20:$ \\
O\,II (26)   &      4395.95   &              &           \\
C\,II (40)   &      4409.98   &     0.97     &    $-16:$ \\
C\,II (39)   &      4411.4:   &      0.96:   &           \\
$[$Fe\,II$]$ F7&    4413.78   & e 1.07       &    $-51 $ \\
O\,II (5)    &      4414.91   &     0.84     &    $-19 $ \\
$[$Fe\,II$]$ F6&    4416.27   &  e 1.05      &    $-50 $ \\
O\,II (5)    &      4416.98   &      0.89    &    $-21 $ \\
Fe\,III (4)  &      4419.60   &   e 1.04     &    $-57 $ \\
             &                &     0.93     &    $-19:$ \\
Fe\,III (4)  &      4430.95   &      0.96:   &           \\
N\,II (55)   &      4432.74   &      0.97:   &    $-28:$ \\
He\,I (50)   &      4437.55   &      0.87    &    $-23 $ \\
N\,II (15)   &      4447.03   &      0.95    &    $-25 $ \\
O\,II (35)   &      4448.21   &      0.98:   &    $-26:$ \\         
$[$Fe\,II$]$ F7&    4452.11   &   e 1.03     &    $-52:$ \\
O\,II (5)    &      4452.38   &      0.96    &    $-25 $ \\
$[$Fe\,II$]$ F6 &   4457.95   &   e 1.03     &    $-49 $ \\
He\,I (14)   &      4471.52   &   e 1.06     &    $-77 $ \\
             &                &  e 0.74      &    $-32 $ \\
             &                &    0.70      &    $-26 $ \\       
$[$Fe\,II$]$ F7&    4474.91   &   e 1.02:    &    $-50:$ \\
Al\,III (8)  &      4479.93   &      0.97:   &    $-26:$ \\
Mg\,II (2)   &      4481.22   &   e 1.04     &    $-72 $ \\
             &                &     0.76     &    $-28 $ \\
Al\,III (8)  &      4512.56   &       0.98   &    $-29:$ \\
Al\,III (8)  &      4529.1:   &        0.96: &           \\      
N\,II (59)   &      4530.41   &       0.98:  &           \\                  
Si\,III (2)  &      4552.62   &       0.75   &    $-19 $ \\                   
Si\,III (2)  &      4567.82   &       0.79   &    $-18 $ \\                 
Si\,III (2)  &      4574.76   &       0.88   &    $-21 $ \\
O\,II (15)   &       4590.97  &        0.90  &    $-50 $  \\
O\,II (15)   &       4596.17  &        0.92: &    $-52 $  \\
N\,II (5)    &       4601.48  &        0.94  &    $-34:$  \\
O\,II (93)   &       4602.11  &              &            \\
N\,II (5)    &       4607.15  &        0.95: &    $-37 $  \\
O\,II (93)   &       4609.42  &        0.96: &    $-49:$  \\
N\,II (5)    &       4613.87  &         0.96:&    $-39:$  \\
C\,II (50)   &       4619.23  &         0.97:&    $-47:$  \\            
N\,II (5)    &       4621.39  &         0.97:&    $-25:$  \\                                       
N\,II (5)    &       4630.54  &         0.88 &    $-37 $  \\       
O\,II (1)    &       4638.85  &         0.89 &    $-42 $  \\
O\,II (1)    &       4641.81  &         0.83 &    $-42 $  \\
N\,II (5)    &       4643.09  &         0.93 &    $-35:$  \\
C\,III (1)   &       4647.42  &         0.94 &    $-44 $  \\                 
O\,II (1)    &       4649.14  &         0.78 &    $-43 $  \\                  
O\,II (1)    &       4650.84  &         0.87 &    $-46:$  \\      
O\,II (1)    &       4661.64  &         0.87 &    $-42 $  \\       
O\,II (1)    &       4673.75  &         0.96 &    $-45 $  \\ 
O\,II (1)    &       4676.23  &         0.89 &    $-42 $  \\
N\,II (62)   &       4678.14  &         0.98:&    $-47:$  \\
Si\,III (13) &       4683.02  &         0.99:&            \\
Si\,III (13) &       4683.8   &              &            \\
O\,II (1)    &       4696.36  &         0.98:&    $-44:$  \\
O\,II (25)   &       4699.2:  &         0.94 &    $-50:$  \\
O\,II (25)   &       4705.36  &         0.92:&    $-42:$  \\
O\,II (24)   &       4710.01  &         0.97:&    $-46:$  \\
He\,I (12)   &       4713.18  &      e 1.00: &    $-75:$  \\                
             &                &         0.78 &    $-27 $  \\
Si\,II       &       4716.65  &         0.99:&    $-50:$  \\
N\,II (20)   &       4779.72  &         0.99:&    $-34:$  \\
N\,II (20)   &       4788.13  &         0.98:&    $-45:$  \\
N\,II (20)   &       4803.29  &         0.98:&    $-48:$  \\
Ar\,II (6)   &       4806.02  &         0.98:&    $-43:$  \\
Si\,III (9)  &       4813.33  &         0.96:&    $-38:$  \\       
$[$Fe\,II$]$ F20&    4814.55  &      e 1.06  &    $-50 $  \\
Si\,III (9)  &       4819.72  &         0.96 &    $-46 $  \\
Si\,III (9)  &       4828.97  &         0.97 &    $-46:$  \\
H$\beta$     &       4861.33  &         1.23 &    $-75 $  \\
                              &      e 2.78  &    $-33 $  \\
O\,II (57)   &       4871.52  &        0.98: &    $-43:$  \\
$[$Fe\,II$]$ F4&     4889.63  &      e 1.03  &    $-49 $  \\
O\,II (28)   &       4890.86  &        0.98: &    $-44:$  \\
$[$Fe\,II$]$ F20&    4905.35  &      e 1.01: &            \\            
O\,II (28)    &      4906.83  &        0.96: &    $-50:$  \\
S\,II (15)    &      4917.21  &       0.98:  &    $-49:$  \\
He\,I (48)    &      4921.93  &       0.63   &    $-29 $  \\
S\,II (7)     &      4924.12  &       0.93:  &            \\
O\,II (28)    &      4924.53  &              &            \\
S\,II (7)     &      4925.35  &       0.98:  &    $-41:$  \\
O\,II (33)    &      4941.07  &       0.98   &    $-41:$  \\
O\,II (33)    &      4943.00  &       0.97   &    $-45 $  \\
N\,II (24)    &      4994.36  &       0.97   &    $-41:$  \\
N\,II (19)    &      5001.35: &       0.90   &    $-47 $  \\
N\,II (4)     &      5002.70  &       0.97:  &    $-44:$  \\
N\,II (19)    &      5005.15  &       0.91   &    $-45 $  \\
N\,II (24)    &      5007.33  &       0.98   &    $-43:$  \\
N\,II (4)     &      5010.62  &     0.96     &    $-42 $  \\
S\,II (15)    &      5014.07  &        0.97: &    $-43:$  \\
He\,I (4)     &      5015.68  &    e 1.14    &    $-65 $  \\
              &               &       0.76   &    $-23 $  \\
N\,II (19)    &      5025.66  &       0.99:  &    $-42:$  \\
S\,II (1)     &      5027.22  &       0.98:  &    $-48:$  \\
S\,II (7)     &      5032.45  &       0.99:  &    $-49:$  \\
Si\,II (5)    &      5041.03  &    e 1.05    &    $-53 $  \\
N\,II (4)     &      5045.10  &       0.93   &    $-44 $  \\
He\,I (47)    &      5047.74  &       0.83   &    $-34 $  \\
Si\,II (5)    &      5055.97  &    e 1.12    &    $-55 $  \\
Fe\,III (5)   &      5073.90  &       0.96   &    $-28:$  \\
Fe\,III (5)   &      5086.72  &    e 1.00:   &    $-66:$  \\
              &               &      0.97    &    $-27:$  \\
Fe\,III (5)   &      5127.5:  &      e 1.04  &    $-72:$  \\
              &               &       0.94   &    $-27 $  \\
C\,II (16)    &      5133.1:  &         0.92 &    $-47 $  \\
C\,II (16)    &      5139.17  &        0.98: &    $-47:$  \\
C\,II (16)    &      5143.49  &        0.96  &    $-46 $  \\
C\,II (16)    &      5145.16  &        0.92  &    $-44:$  \\
Ar\,II (13)   &      5145.31  &              &            \\
C\,II (16)    &      5151.09  &       0.96   &    $-54:$  \\
Fe\,III (5)   &      5156.11  &    e 1.04    &    $-69:$  \\
              &               &      0.91    &    $-31 $  \\
$[$Fe\,II$]$ F19&    5158.78  &     e 1.11   &    $-52 $  \\
O\,II (32)    &      5160.02  &        0.98: &    $-47:$  \\
Fe\,III (5)   &      5193.91  &     e 1.01:  &    $-64:$  \\
              &               &      0.98    &    $-30:$  \\
$[$N\,1$]$ F1 &      5197.90  &      e 1.17  &    $-50 $  \\
$[$N\,1$]$ F1 &      5200.26  &      e 1.09  &    $-51 $  \\
O\,II (32)    &      5206.65  &         0.99 &    $-46 $  \\
S\,III        &      5219.32  &         0.99:&            \\
$[$Fe\,II$]$ F19&    5261.62  &      e 1.04  &    $-51 $  \\
$[$Fe\,II$]$ F18&    5273.35  &      e 1.03: &    $-50 $  \\
O\,I (26)       &    5299.04  &      e 1.04  &    $-51 $  \\
$[$Fe\,II$]$ F19&    5376.47  &      e 1.02: &    $-48:$  \\
S\,II (6)       &    5428.67  &         0.99:&    $-40 $  \\               
S\,II (6)       &    5432.82  &         0.98:&    $-44:$  \\
S\,II (6)       &    5453.83  &              &            \\
N\,II (29)      &    5454.22  &        0.92  &    $-48:$  \\
\hline            
                 \multicolumn{4}{c}{\underline{18.01.05}} \\
N\,II (29)    &      5495.67  &         0.97 &     $-59  $ \\                
S\,II (6)     &      5606.15  &         0.98:&     $-59: $ \\               
S\,II (14)    &      5639.97  &         0.93 &     $-60: $ \\               
S\,II (11)    &      5640.33  &              &             \\
C\,II (15)    &      5640.55  &              &             \\
S\,II (14)    &      5647.03  &          0.97:&            \\           
S\,II (11)    &      5659.99  &          0.99:&    $-58: $ \\            
C\,II (15)    &      5662.47  &          0.97 &    $-60: $ \\           
N\,II (3)     &      5666.63  &          0.90 &    $-53  $ \\            
N\,II (3)     &      5676.02  &          0.90 &    $-52: $ \\              
N\,II (3)     &      5679.56  &          0.84 &    $-53  $ \\        
N\,II (3)     &      5686.21  &          0.95 &    $-57: $ \\       
Al\,III (2)   &      5696.60  &          0.85 &    $-56  $ \\      
N\,II (3)     &      5710.77  &          0.94:&    $-60: $ \\         
Al\,III (2)   &      5722.73  &          0.90:&    $-53: $ \\          
Si\,III (4)   &      5739.73  &          0.84 &    $-53  $ \\        
DIB           &      5780.37  &          0.96 &    $-14: $ \\         
DIB           &      5796.96  &          0.97:&    $-16: $ \\      
Fe\,III (114) &      5833.94  &          0.97:&    $-53: $ \\     
He\,I (11)    &      5875.72  &          0.90 &    $-70: $ \\
              &               &        e 1.10 &    $-30  $ \\            
Na\,I (1)     &      5889.95  &          0.32 &    $-72.4$ \\           
              &               &          0.39 &    $-65.3$ \\
              &               &          0.60 &    $-52.8$ \\          
              &               &          0.40 &    $-27.5$ \\           
              &               &          0.05 &    $-10.0$ \\          
Na\,I (1)     &      5895.92  &          0.49 &    $-72.5$ \\         
              &               &          0.57 &    $-64.7$ \\
              &               &          0.74 &    $-52.1$ \\                                                             
              &               &          0.55 &    $-27.7$ \\          
              &               &          0.06 &    $ -9.7$ \\        
C\,II (5)     &       5891.59 &        e 1.07 &    $-55: $ \\
              &               &          0.96 &    $-18: $ \\  
Si\,II (4)    &       5957.56 &        e 1.08 &    $-52  $ \\              
O\,I (23)     &       5958.5: &        e 1.05 &    $-52: $ \\              
Si\,II (4)    &       5978.93 &        e 1.18 &    $-51  $ \\            
Fe\,III (117) &       6032.59 &        e 1.03:&    $-59: $ \\
O\,I (22)     &       6046.39:&        e 1.12 &    $-52: $ \\
Ne\,I (3)     &       6074.34 &          0.98:&    $-45: $ \\
C\,II (24)    &       6095.29 &        e 1.02:&    $-53: $ \\
C\,II (24)    &       6098.51 &        e 1.02:&    $-53: $ \\
Ne\,I (1)     &       6143.06 &          0.97:&    $-64: $ \\
DIB           &       6195.96 &          0.96 &    $-13  $ \\
$[$O\,I$]$ F1 &       6300.30 &       e 1.07: &    $-48: $ \\
Si\,II (2)    &       6347.10 &        e 1.15 &    $-50  $ \\
Si\,II (2)    &       6371.36 &        e 1.08 &    $-49  $ \\
DIB           &       6379.29 &         0.93: &            \\
Ne\,I (1)     &       6402.25 &         0.92: &    $-63: $ \\
$[$N\,II$]$ F1&       6548.03 &       e 1.13  &    $-51  $ \\
H$\alpha$     &       6562.81 &         2.06  &    $-78  $ \\
              &               &       e 5.90  &    $-49  $ \\
              &               &       e 4.48: &    $-24  $ \\
C\,II (2)     &       6578.05 &           0.70&    $-51  $ \\
C\,II (2)     &       6582.88 &          0.70:&    $-55: $ \\
$[$N\,II$]$ F1&       6583.45 &         e 1.1:&    $-51: $ \\
DIB           &       6613.56 &          0.95 &    $-10.8$ \\
O\,II (4)     &       6641.05 &          0.97 &    $-53: $ \\
$[$S II$]$ F2 &       6716.47 &         1.03: &    $-53: $ \\
O\,II (4)     &       6721.35 &         0.97: &    $-55: $ \\
$[$S\,II$]$ F2&       6730.85 &         1.09  &    $-53  $ \\                                    
\label{lines}                                 
\end{longtable}

\end{document}